\documentstyle[11pt,amssymb,epsf,fleqn]{article}
\begin{document}
\title{Neutral Higgs sector of the next-to-minimal supersymmetric standard model with explicit CP violation}
\author{S.W. Ham
\\
{\it Center for High Energy Physics, Kyungpook National University Taegu 702-701, Korea}
\\
\\
S.K. Oh
\\
{\it Center for High Energy Physics, Kyungpook National University Taegu 702-701, Korea}
\\
{\it and Department of Physics, Konkuk University, Seoul 143-701, Korea}
\\
\\
D. Son
\\
{\it Center for High Energy Physics, Kyungpook National University Taegu 702-701, Korea}
\\
\\}
\date{}
\maketitle
\begin{abstract}
The neutral Higgs sector of the next-to-minimal supersymmetric standard model (NMSSM) with explicit CP violation is investigated at the 1-loop level, using the effective potential method; not only the loops involving the third generation of quarks and scalar quarks, but also the loops involving $W$ boson, charged Higgs boson, and chargino are taken into account.
It is found that for some parameter values of the NMSSM the contributions from the $W$ boson, charged Higgs boson, and chargino loops may modify the masses of the neutral Higgs bosons and the mixings among them significantly, depending on the CP phase.
In $e^+e^-$ collisions, the prospects for discovering neutral Higgs bosons are investigated within the context of the NMSSM with explicit CP violation when the dominant component of the lightest neutral Higgs boson is the Higgs singlet field of the NMSSM.
\end{abstract}
\vfil

\section{INTRODUCTION}

In the standard model, the phenomenology of CP violation is described in terms of a complex phase in the Cabibbo-Kobayashi-Maskawa (CKM) mixing matrix for quarks [1].
The source of the complex phase is not clarified in the standard model.
If there are two Higgs doublets in the model, the complex phase might originate from the Higgs sector.
As the supersymmetrically extended versions of the standard model require at least two Higgs doublets [2], to give separate masses for the up-type quarks and the down-type quarks, they are regarded as natural candidates for explaining the origin of CP violation.
This is all the more because supersymmetry is broken in nature at the electroweak scale.
One would obtain no CP phase in the Higgs sector if supersymmetry is intact.

In supersymmetric standard models, the CP violation can occur in the neutral Higgs sector through the mixing between the scalar and pseudoscalar Higgs bosons.
In principle, it can be accomplished either spontaneously or explicitly. 
Either way, the phenomenology of the CP violation in the supersymmetric standard models is considered as particularly useful probes of new physics at the high energy scale because CP violation only appears through the CKM matrix in the standard model. 
Recently, the attempts to explain the source of the CP violation within the context of supersymmetry have been investigated by many authors [3-5].

In the minimal supersymmetric standard model (MSSM), many authors have investigated if it is possible to violate the CP symmetry either spontaneously [3] or explicitly [4,5].
It is found that the tree level Higgs potential of the MSSM is unable to break the CP symmetry in either way. 
Further investigations at the 1-loop level of the MSSM show that the scenario of spontaneous CP violation is experimentally ruled out by the Higgs search at the CERN $e^+ e^-$ collider LEP, because it leads to a very light neutral Higgs boson [3].
However, at the 1-loop level, the scenario of explicit CP violation in the MSSM
is found to be viable through the radiatively corrected Higgs potential. 
Subsequently, a number of investigations have been devoted to the effects of explicit CP violation in the MSSM at the 1-loop level to the charged Higgs boson as well as to the neutral ones by considering radiative corrections due to the third generation quark and scalar quark contributions [4].
Recently the MSSM searches for explicit CP violation were extended to radiative corrections due to the $W$ boson, the charged Higgs boson, and to the chargino contribution [5].

Meanwhile, the CP symmetry may also be broken if the Higgs sector is extended by introducing at least one Higgs singlet.
The supersymmetric standard models with additional neutral Higgs singlets can evade some of the strong constraints of the MSSM.
These models may be classified according to whether they have a discrete symmetry or not.
It is known that for those models without a discrete $Z_3$ symmetry, spontaneous CP violation is possible at the tree level [6] and up to the 1-loop level including radiative corrections due to the top and scalar top quark loops [7].
On the other hand, those models with discrete symmetry can cause a domain problem associated with the existence of degenerate vacua after the electroweak phase transition [8].
The possible domain wall problem is assumed to be solved by adding nonrenormalizable interactions which break the $Z_3$ symmetry without spoiling the quantum stability [9].
Among those with a discrete $Z_3$ symmetry on its Higgs potential, the simplest model, with just one Higgs singlet, is usually called as the next-to-minimal supersymmetric standard model (NMSSM) [10].
The NMSSM is basically different from the MSSM in that it has one extra Higgs singlet.

For spontaneous CP violation, it is well known that the NMSSM, which has $Z_3$ symmetry in its Higgs potential cannot produce spontaneous CP violation at the tree level, because of vacuum stability [11].
This behavior does not change by taking into account the contributions of the degenerate scalar top quark loops [12].
Spontaneous CP violation can only occur if the scalar top quark masses at the 1-loop effective potential are nondegenerate [13].
However, at the tree level, unlike the MSSM, the scenario of explicit CP violation is possible for the NMSSM.
By assuming the degeneracy of the top scalar quark masses in the Higgs sector of the NMSSM, it is found that large explicit CP violation may be realized as the vacuum expectation value (VEV) of the neutral Higgs singlet approaches the electroweak scale [14].

In a previous article [15], we have analyzed the phenomenology of explicit CP violation in the Higgs sector of the NMSSM at the 1-loop level.
Using the effective potential method, we have included the radiative corrections due to quarks and scalar quarks of the third generation.
We have found that there are parameter regions of the NMSSM at the 1-loop level where the lightest neutral Higgs boson may even be massless, without being detected at LEP2.
The VEV of the neutral Higgs singlet has been shown to be no smaller than 16 GeV for any parameter values of the NMSSM with explicit CP violation.
This value of the lower limit is found to increase up to about 45 GeV as the ratio ($\tan \beta$) of the VEVs of the two Higgs doublets decreases to smaller values ($\sim$ 2).
The discovery limit of the Higgs boson search at LEP2 is found to cover about one half of the kinematically allowed part of the whole parameter space of the NMSSM, and the portion is roughly stable against the CP phase.

In this article, we continue our analysis more exhaustively by including the contributions from the loops involving the $W$ boson, charged Higgs boson, and chargino, in addition to those from the loops involving quarks and scalar quarks of the third generation.
It is found that the inclusion of the $W$ boson, charged Higgs boson, and chargino loops yields significant modifications to the results on the neutral Higgs sector obtained without them, depending on the parameter values of the model.
At the 1-loop level two CP phases arise from each scalar quark mass matrix for the third generation.
We assume that at the tree level the CP phase is equal to two CP ones arising from the contribution of radiative corrections due to the third generation quark and scalar quarks.
One CP phase appears from the 1-loop effective potential coming from the chargino sector.
We assume that the CP phase arising from the chargino sector does not equal the other ones at the 1-loop level.
We investigate the size of the contribution of the $W$ boson, the charged Higgs boson, and the charginos to the neutral Higgs boson mass at the 1-loop level.
We then investigate the phenomenological consequences for the neutral Higgs sector of the NMSSM with explicit CP violation at future $e^+e^-$ collisions.

\section{EFFECTIVE POTENTIAL}

The additional Higgs singlet superfield in the NMSSM can avoid the so-called $\mu$-parameter problem in the MSSM, where the dimensional $\mu$ parameter of the MSSM is introduced by hand.
In the NMSSM, the corresponding quantity is generated dynamically.
The additional Higgs singlet superfiled, $N$, has a vanishing hypercharge and neutral.
The neutrality of $N$ enlarges the spectrum of neutral particles of the NMSSM with an additional scalar Higgs boson, an additional pseudoscalar Higgs boson, and an additional neutralino, while the structure of the charged sector of the NMSSM is essentially the same as that of the MSSM.

Ignoring all quark and lepton Yukawa couplings except for those of the third generation quarks, the superpotential involving the superfields $H_1^T$ = ($H_1^0$,$H_1^-$), $H_2^T$ = ($H_2^+$,$H_2^0$), and $N$ in the NMSSM with discrete $Z_3$ symmetry is given by
\[
W = h_b H_1^T \epsilon Q b_R^c - h_t H_2^T \epsilon Q t_R^c
+ \lambda H_1^T \epsilon H_2 N - {k \over 3} N^3 \ ,
\]
where $h_t$ and $h_b$ are Yukawa coupling constants of the quarks for the third generation, and the parameters $\lambda$ and $k$ are dimensionless coupling constants.
The 2 $\times$ 2 antisymmetric matrix $\epsilon$ is defined as $\epsilon_{12}$ = $ - \epsilon_{21}$ = 1.
The SU(2) doublet superfield $Q^T$ = ($t_L$,$b_L$) contains the left-handed quark for the third generation, and the SU(2) singlet superfields $t_R^c$ and $b_R^c$ are the charge conjugates of the right-handed quarks for the third generation. 
At the tree level, a massless pseudoscalar Higgs boson would appear if the global U(1) Peccei-Quinn symmetry would not be broken explicitly by the cubic term in the Higgs singlet in the superpotential.

In the NMSSM one breaks the supersymmetry by introducing explicitly soft breaking terms to the Lagrangian.
The relevant soft breaking terms for the above superpotential can be written as
\begin{eqnarray}
{\cal L}_{\rm soft} & = &\mbox{} - m_{H_1}^2|H_1|^2
- m_{H_2}^2|H_2|^2 - m_N^2|N|^2 - m_Q^2 |{\tilde Q}|^2
- m_T^2 |{\tilde t}_R^c|^2 - m_B^2 |{\tilde b}_R^c|^2 \cr
& &\mbox{} + ( - h_b A_b H_1^T \epsilon {\tilde Q} {\tilde b}_R^c
+ h_t A_t H_2^T \epsilon {\tilde Q} {\tilde t}_R^c + {\rm H.c.}) \cr
& &\mbox{} + \left (\lambda A_{\lambda} H_1^T \epsilon H_2 N
+ {k \over 3} A_k N^3 + {\rm H.c.}\right )
+ \left ({1 \over 2} M_2 \lambda_2^a \lambda_2^a
+ {1 \over 2} M_1 \lambda_1 \lambda_1 + {\rm H.c.} \right) \ ,
\end{eqnarray}
where $m_{H_1}^2$, $m_{H_2}^2$, $m_N^2$, $m_Q^2$, $m_T^2$, and $m_B^2$ are the soft breaking masses; $A_{\lambda}$, $A_k$, $A_t$, and $A_b$ are the trilinear soft SUSY breaking parameters with mass dimension; $\lambda_2^a$ $(a = 1,2,3)$ and $\lambda_1$ are the gauginos of the SU(2) and U(1) gauge groups; and $M_2$ and $M_1$ are their mass parameters, respectively.
The full tree level Higgs potential $V^0$ can then be obtained by collecting the relevant $F$ terms and $D$ terms in the superpotential, to add to the Higgs part of the above soft breaking terms.

The condition for the electroweak symmetry breaking is obtained by minimizing the Higgs potential $V = V^0$ with respect to the VEVs of the three Higgs fields.
We assume that the VEVs of three neutral Higgs fields $H_1^0$, $H_2^0$, and $N$ are, respectively, $v_1$, $v_2$, and $x$, all being real and positive, so that there is no spontaneous CP violation in the Higgs sector in the NMSSM [11-13].
The ratio of the two VEVs is denoted by $\tan \beta$ = $v_2/v_1$, giving the $Z$ boson mass as $m_Z^2$ = $(g_1^2 + g_2^2) v^2/2$, and the $W$ boson mass as $m^2_W = g^2_2 v^2/2$, with $v = \sqrt{v^2_1 + v^2_2} =$ 175 GeV.
The soft SUSY breaking masses $m_{H_1}^2$, $m_{H_2}^2$, and $m_N^2$ can be eliminated by imposing the minimum conditions on the Higgs potential.

The explicit CP violation in the Higgs sector of the NMSSM might be activated if there are complex phases in the Higgs potential.
From the tree level Higgs potential of the NMSSM, complex phases might occur by assuming that $\lambda$, $A_{\lambda}$, $k$, and $A_k$ are complex numbers. 
Among them, $\lambda A_{\lambda}$ and $k A_k$ can be adjusted to be real and positive by redefining the overall phases of the superfields $H_1$, $H_2$, and $N$.
Then, we are left essentially with one phase: It may be chosen to be the one in $\lambda k^*$ = $\lambda k e^{i \phi_0}$, which allows for the explicit CP violation at the tree level. 
This phase appears in the expression for the mass of the charged Higgs boson [15]:
\begin{equation}
m_{C^+}^2 = m_W^2 - \lambda^2 v^2 + \lambda (A_{\lambda} + k x \cos \phi_0)
{2 x \over \sin 2 \beta} \ .
\end{equation}
Note that although the phase $\phi_0$ is present in the potential, it does not show up in the vacuum.
Thus, one need not impose a minimum condition on the Higgs potential with respect to it.
Consequently, the tree level Higgs potential contains $\phi_0$, $\tan \beta$, $\lambda$, $A_{\lambda}$, $k$, $A_k$, and $x$ as free parameters.

Now, at the 1-loop level, in the NMSSM with explicit CP violation, the radiative corrections to the tree level Higgs potential come primarily from the quark and scalar quark loops of the third generation: Generally, the leading radiative corrections come from the top and scalar top quark loops.
For large $\tan \beta$, the contributions of the bottom and scalar bottom quark loops can be significantly large.
Nevertheless, in order to carry out a comprehensive study on the phenomenology of the explicit CP violation at the 1-loop level in the NMSSM, one should also include the contributions of the $W$ boson, the charged Higgs boson, and the charginos loops.
Let us decompose radiative corrections to the 1-loop effective potential into three parts: a part for the radiative corrections from the top and scalar top quark loops, another one for those from the bottom and scalar bottom quark loops, and the third one for the radiative corrections from the $W$ boson, the charged Higgs boson, and the charginos.
Thus, the radiative corrections to the 1-loop effective Higgs potential is written as [16]
\begin{equation}
V^1 = V^t + V^b + V^{\chi} \ ,
\end{equation}
with
\begin{eqnarray}
V^t & = & {3 \over 32\pi^2} \left \{ \sum_{i = 1}^2 {\cal M}_{\tilde{t_i}}^4 \left (\log {{\cal M}_{\tilde{t_i}}^2 \over \Lambda^2} - {3\over 2} \right )
   - 2 {\cal M}_t^4 \left (\log {{\cal M}_t^2 \over \Lambda^2}
   - {3\over 2} \right ) \right \}                      \ ,     \cr
V^b & = & {3 \over 32\pi^2} \left \{ \sum_{i = 1}^2 {\cal M}_{\tilde{b_i}}^4 \left (\log {{\cal M}_{\tilde{b_i}}^2 \over \Lambda^2} - {3\over 2} \right )
   - 2 {\cal M}_b^4 \left (\log {{\cal M}_b^2 \over \Lambda^2}
   - {3\over 2} \right ) \right \}                       \ ,    \cr
V^{\chi} & = & {1 \over 32\pi^2} \left \{
3 {\cal M}_W^4 \left (\log {{\cal M}_W^2 \over \Lambda^2}
   - {3\over 2} \right )
+ {\cal M}_{C^+}^4 \left (\log {{\cal M}_{C^+}^2 \over \Lambda^2}
   - {3\over 2} \right ) \right \} \cr
    &   &\mbox{} - {1 \over 16 \pi^2}\left \{ \sum_{i = 1}^2 {\cal M}_{{\tilde \chi}_i}^4 \left (\log {{\cal M}_{{\tilde \chi}_i}^2 \over \Lambda^2} - {3\over 2}  \right )  \right \}    \ ,
\end{eqnarray}
where ${\cal M}$ denote the mass matrices for the particles and superparticles with the field dependence, and $\Lambda$ is the renormalization scale in the modified minimal substraction (${\overline {\rm MS}}$) scheme.
The full 1-loop level Higgs potential will be given by adding $V^1$ to the tree level Higgs potential.

Among the mass matrices, there is a 4 $\times$ 4 Hermitian mass matrix for the scalar quarks of the third generation, which may be broken up into four 2 $\times$ 2 block submatrices.
After the electroweak symmetry breaking, only two diagonal block submatrices become dependent on the neutral Higgs fields and thus have nonzero eigenvalues.
By diagonalizing each diagonal block submatrix, one can obtain the 1-loop masses of the scalar top quarks and the scalar bottom quarks separately.
From the complexity of the relevant parameters, a phase appears in the expressions for the masses of the scalar top quarks and another one for the masses of the scalar bottom quarks.
These two phases may also take part in the explicit CP violation at the 1-loop level.
In general, the two phases are different.
However, for simplicity, we assume that $A_b$ = $A_t$, while keeping them complex.
Then, the two phases become equal.

The masses of the scalar top quarks $m_{{\tilde t}_1}^2$ and $m_{{\tilde t}_2}^2$ are obtained as
\begin{equation}
  m_t^2 + {\textstyle {1 \over 2}} (m_Q^2 + m_T^2) \mp
\sqrt{{\textstyle {1 \over 4}} (m_Q^2 - m_T^2)^2 + m_t^2 (A_t^2 + \lambda^2 x^2 \cot^2 \beta
+ 2 A_t \lambda x \cot \beta \cos \phi)} \ , \nonumber
\end{equation}
and those of the scalar bottom quarks $m_{{\tilde b}_1}^2$ and $m_{{\tilde b}_2}^2$ as
\begin{equation}
 m_b^2 + {\textstyle {1 \over 2}} (m_Q^2 + m_T^2) \mp
\sqrt{{\textstyle {1 \over 4}} (m_Q^2 - m_T^2)^2 + m_b^2 (A_t^2 + \lambda^2 x^2 \tan^2 \beta
+ 2 A_t \lambda x \tan \beta \cos \phi)} \ , \nonumber
\end{equation}
where the contribution from the $D$ terms are not included.
Thus the CP phase in the bottom scalar quark masses is equal to that of the top scalar ones: $\phi$ is the relative phase between $A_t = A_b$ and $\lambda$, allowing for the explicit CP violation in the Higgs sector through radiative corrections.
The phase $\phi$ would be absent as long as both the scalar top quarks and scalar bottom quarks are degenerate in masses.
In the scalar quark masses $\phi$ has no CP-violating effects on the Higgs sector, since it could be replaced by some modification of $A_t$ anyhow.
However, unless $\phi$ vanishes, explicit CP violation occurs in the Higgs couplings through radiative corrections.
We assume that $\phi$ is equal to $\phi_0$, in the mass of the charged Higgs boson, for simplicity.
Note that the top quark mass is given by $m_t^2$ = $h_t^2 v^2 \sin^2 \beta$, and the bottom quark mass as $m_b^2$ = $h_b^2 v^2 \cos^2 \beta$.

The square masses of charginos are given by the eigenvalues of the square-mass matrix
\[
\begin{array}{ccc}
{\cal M}_{\tilde \chi}^2 & = & \left ( \begin{array}{cc}
  |M_2|^2 + g_2^2 |H_1|^2   & g_2 (M_2 H_2^{0 *} - \lambda^* H_1^0 N^*)   \cr
g_2 (M_2^* H_2^{0} - \lambda^* H_1^{0 *} N)  & |\lambda|^2 |N|^2
  \end{array} \right )    \cr
\end{array}
\]
as 
\begin{eqnarray}
m_{{\tilde \chi}_{1, \ 2}}^2 & \simeq &
{\textstyle {1 \over 2}} \{M_2^2 + \lambda^2 x^2 + 2 m_W^2 \}
\mp [ {\textstyle {1 \over 4}} \{M_2^2 - \lambda^2 x^2 \}^2
+ m_W^2 \{M_2^2 - \lambda^2 x^2\} \cos 2 \beta \cr
& &\mbox{} + 2 m_W^2 \{(M_2 \sin \beta)^2
+ (\lambda x \cos \beta)^2 - M_2 \lambda x \sin 2 \beta \cos \phi_c \}
]^{1 / 2}      \ ,
\end{eqnarray}
where we neglect the quartic term of $g_2$ since its order is equal to that of the $D$ terms in the scalar quark sector.
Notice that there is a complex phase in the chargino masses, $\phi_c$, arising from the relative phase between the SU(2) gaugino mass $M_2$ and $\lambda$.
We assume that $\phi_c$ is not equal to $\phi$.
The total Higgs potential at the 1-loop level is written as $V$ = $V^0 + V^1$.
Here, $V^1$ contains $\phi_c$, $m_Q$, $m_T$, $A_t$, and $M_2$ as free parameters.

The Higgs sector of the NMSSM has ten real degrees of freedom coming from two Higgs doublets and one Higgs singlet.
After the electroweak symmetry breakdown takes place, three of them correspond
to a neutral Goldstone boson and a pair of charged Goldstone bosons, and the other seven correspond to five neutral Higgs bosons and a pair of charged Higgs bosons.
Transforming to a unitary gauge, one may express two Higgs doublets and one Higgs singlet as
\begin{eqnarray}
\begin{array}{ccc}
H_1 & = & \left ( \begin{array}{c}
          v_1 + S_1 + i \sin \beta A   \cr
          \sin \beta C^{+ *}
  \end{array} \right )  \ ,  \cr
H_2 & = & \left ( \begin{array}{c}
          \cos \beta C^+           \cr
          v_2 + S_2 + i \cos \beta A
  \end{array} \right ) \ ,   \cr
N & = & ( \begin{array}{c}
          x + X + i Y
        \end{array} )  \ ,
\end{array}
\end{eqnarray}
where $S_1$, $S_2$, $A$, $X$, and $Y$ are neutral fields and $C^+$ is the charged field.
If CP symmetry is conserved in the Higgs potential, $S_1$, $S_2$, and $X$ would become the CP even scalar Higgs fields, whereas $A$ and $Y$ would become the CP odd pseudoscalar Higgs fields.
The three orthogonal combinations among $A$ and $C^{\pm}$ yield one neutral Goldstone boson and a pair of charged Goldstone bosons; they will eventually be absorbed into the longitudinal component of $Z$ and $W$ gauge bosons, respectively.

\section{NEUTRAL HIGGS BOSON MASSES}

In the NMSSM with explicit CP violation the neutral Higgs boson mass matrix is given as a $5 \times 5$ matrix that is obtained by the second derivatives of the Higgs potential with respect to the five neutral Higgs fields.
On the basis of ($S_1$,$S_2$,$A$,$X$,$Y$) the elements of the symmetric mass matrix for the neutral Higgs boson can be expressed as
\begin{equation}
M_{ij} = M_{ij}^0 + \delta M_{ij}^t + \delta M_{ij}^b + \delta M_{ij}^{\chi}
\ (i,j = 1 \ {\rm to} \ 5) \ ,
\end{equation}
where we note that $M_{ij}^0$ comes from the tree level Higgs potential, $\delta M_{ij}^t$ from the contribution of the top and scalar top quark loops, $\delta M_{ij}^b$ from the contribution of the bottom and scalar bottom quark loops, and $\delta M_{ij}^{\chi}$ from the loops of the $W$ boson, the charged Higgs boson, and the charginos. 
Thus, $M_{ij}^0$ is obtained from $V^0$, $\delta M_{ij}^t$ from $V^t$, $\delta M_{ij}^b$ from $V^b$, and $\delta M_{ij}^{\chi}$ from $V^{\chi}$.

The complicated expressions for $M_{ij}^0$, $\delta M_{ij}^t$, and $\delta M_{ij}^b$ are given elsewhere [15]. 
We concentrate on $\delta M_{ij}^{\chi}$.
We calculate each elements of it. 
The results read
\begin{eqnarray}
\delta M_{11}^{\chi} & = & \mbox{} - {m_W^4 \cos^2 \beta \over 4 \pi^2 v^2}
{\Delta _{{\tilde \chi}_1}^2 g(m_{{\tilde \chi}_1}^2, \ m_{{\tilde \chi}_2}^2)
\over (m_{{\tilde \chi}_2}^2 - m_{{\tilde \chi}_1}^2)^2}
+ {m_W^2 \over 4 \pi^2 v^2}
M_2 \lambda x \tan \beta \cos \phi_c
f(m_{{\tilde \chi}_1}^2, \ m_{{\tilde \chi}_2}^2) \cr
& &\mbox{} - {m_W^4 \cos^2 \beta \over 2 \pi^2 v^2}
{\Delta_{{\tilde \chi}_1} \log ({m_{{\tilde \chi}_2}^2 / m_{{\tilde \chi}_1}^2}) \over (m_{{\tilde \chi}_2}^2 - m_{{\tilde \chi}_1}^2)}
+ {m_W^4 \cos^2 \beta \over 8 \pi^2 v^2} \log \left ({m_W^6 m_{C^+}^2 \over
m_{{\tilde \chi}_1}^4 m_{{\tilde \chi}_2}^4} \right ) \cr
& &\mbox{} + {\sin^2 \beta \over 16 \pi^2 v^2} (2 \lambda^2 v^2 - m_W^2)
\left \{m_{C^+}^2 \log \left ({m_{C^+}^2 \over \Lambda^2} \right ) - m_{C^+}^2 \right \}  \ , \cr
 & & \cr
\delta M_{22}^{\chi} & = & \mbox{} - {m_W^4 \sin^2 \beta \over 4 \pi^2 v^2}
{\Delta _{{\tilde \chi}_2}^2 g(m_{{\tilde \chi}_1}^2, \ m_{{\tilde \chi}_2}^2)
\over (m_{{\tilde \chi}_2}^2 - m_{{\tilde \chi}_1}^2)^2}
+ {m_W^2 \over 4 \pi^2 v^2}
M_2 \lambda x \cot \beta \cos \phi_c
f(m_{{\tilde \chi}_1}^2, \ m_{{\tilde \chi}_2}^2) \cr
& &\mbox{} - {m_W^4 \sin^2 \beta \over 2 \pi^2 v^2}
{\Delta_{{\tilde \chi}_2} \log ({m_{{\tilde \chi}_2}^2 / m_{{\tilde \chi}_1}^2}) \over (m_{{\tilde \chi}_2}^2 - m_{{\tilde \chi}_1}^2)}
+ {m_W^4 \sin^2 \beta \over 8 \pi^2 v^2} \log \left ({m_W^6 m_{C^+}^2 \over
m_{{\tilde \chi}_1}^4 m_{{\tilde \chi}_2}^4} \right ) \cr
& &\mbox{} + {\cos^2 \beta \over 16 \pi^2 v^2} (2 \lambda^2 v^2 - m_W^2)
\left \{m_{C^+}^2 \log \left ({m_{C^+}^2 \over \Lambda^2} \right ) - m_{C^+}^2 \right \}  \ , \cr
 & & \cr
\delta M_{33}^{\chi} & = & \mbox{} - {m_W^4 \over \pi^2 v^2}
{M_2^2 \lambda^2 x^2 \sin^2 \phi_c
\over (m_{{\tilde \chi}_2}^2 - m_{{\tilde \chi}_1}^2)^2}
 g(m_{{\tilde \chi}_1}^2, \ m_{{\tilde \chi}_2}^2)
+ {m_W^2 M_2 \lambda x \cos \phi_c \over \pi^2 v^2 \sin 2 \beta}
f(m_{{\tilde \chi}_1}^2, \ m_{{\tilde \chi}_2}^2) \cr
& &\mbox{} + {(2 \lambda^2 v^2 - m_W^2) \over 16 \pi^2 v^2}
\left \{m_{C^+}^2 \log \left ({m_{C^+}^2 \over \Lambda^2} \right ) - m_{C^+}^2 \right \}
\ , \cr
 & & \cr
\delta M_{44}^{\chi} & = &\mbox{} - {\lambda^2 \over 16 \pi^2}
{\Delta_{\tilde \chi}^2
g(m_{{\tilde \chi}_1}^2, \ m_{{\tilde \chi}_2}^2) \over
(m_{{\tilde \chi}_2}^2 - m_{{\tilde \chi}_1}^2)^2}
+ {m_W^2 \over 8 \pi^2 x} M_2 \lambda \sin 2 \beta \cos \phi_c
f(m_{{\tilde \chi}_1}^2, \ m_{{\tilde \chi}_2}^2) \cr
& &\mbox{} - {\lambda^4 x^2 \over 16 \pi^2}
\log \left ({m_{{\tilde \chi}_1}^2 m_{{\tilde \chi}_2}^2
\over \Lambda^4} \right )
- {\lambda^3 x \over 8 \pi^2}
{\Delta_{\tilde \chi}
\log (m_{{\tilde \chi}_2}^2 / m_{{\tilde \chi}_1}^2)
\over (m_{{\tilde \chi}_2}^2 - m_{{\tilde \chi}_1}^2)} \cr
& &\mbox{} - {\lambda A_{\lambda} \sin 2 \beta \over 32 \pi^2 x}
\left \{m_{C^+}^2 \log \left ({m_{C^+}^2 \over \Lambda^2} \right ) - m_{C^+}^2 \right \} \ , \cr
 & & \cr
\delta M_{55}^{\chi} & = &\mbox{} - {m_W^4 \over 4 \pi^2}
{M_2^2 \lambda^2 \sin^2 2 \beta \sin^2 \phi_c \over
(m_{{\tilde \chi}_2}^2 - m_{{\tilde \chi}_1}^2)^2}
g(m_{{\tilde \chi}_1}^2, \ m_{{\tilde \chi}_2}^2) \cr
& &\mbox{} + {m_W^2 \over 8 \pi^2 x} M_2 \lambda \sin 2 \beta \cos \phi_c
f(m_{{\tilde \chi}_1}^2, \ m_{{\tilde \chi}_2}^2) \cr
& &\mbox{} - {\lambda \sin 2 \beta \over 32 \pi^2 x}
(A_{\lambda} + 8 k x \cos \phi)
\left \{m_{C^+}^2 \log \left ({m_{C^+}^2 \over \Lambda^2} \right ) - m_{C^+}^2 \right \}
 \ ,  \cr
 & & \cr
\delta M_{12}^{\chi} & = &\mbox{} - {m_W^4 \over 8 \pi^2 v^2}
{\sin 2 \beta \Delta_{{\tilde \chi}_1} \Delta_{{\tilde \chi}_2}
\over (m_{{\tilde \chi}_2}^2 - m_{{\tilde \chi}_1}^2)^2}
g(m_{{\tilde \chi}_1}^2, \ m_{{\tilde \chi}_2}^2)
- {m_W^2 \over 4 \pi^2 v^2} M_2 \lambda x \cos \phi_c
f(m_{{\tilde \chi}_1}^2, \ m_{{\tilde \chi}_2}^2) \cr
& &\mbox{} - {m_W^4 \over 8 \pi^2 v^2}
{\sin 2 \beta (\Delta_{{\tilde \chi}_1} + \Delta_{{\tilde \chi}_2})
\over (m_{{\tilde \chi}_2}^2 - m_{{\tilde \chi}_1}^2)}
\log \left ({m_{{\tilde \chi}_2}^2 \over m_{{\tilde \chi}_1}^2} \right) \cr
& &\mbox{} +{m_W^4 \sin 2 \beta \over 16 \pi^2 v^2}
\log \left ( {m_W^6 m_{C^+}^2
\over m_{{\tilde \chi}_1}^4 m_{{\tilde \chi}_2}^4} \right )
- {\lambda^2 m_W^2 \sin 2 \beta \over 8 \pi^2}
\log \left( {m_{C^+}^2 \over \Lambda^2}\right) \cr
& &\mbox{} + {\sin 2 \beta \over 32 \pi^2 v^2}
(m_W^2 - 2 \lambda^2 v^2)
\left \{m_{C^+}^2 \log \left ({m_{C^+}^2 \over \Lambda^2} \right ) - m_{C^+}^2 \right \}  \ , \cr
 & & \cr
\delta M_{13}^{\chi} & = & {m_W^4 \over 2 \pi^2 v^2}
{M_2 \lambda x \cos \beta \sin \phi_c \Delta_{{\tilde \chi}_1} \over
 (m_{{\tilde \chi}_2}^2 - m_{{\tilde \chi}_1}^2)^2}
g(m_{{\tilde \chi}_1}^2, \ m_{{\tilde \chi}_2}^2) \cr
& &\mbox{} - {m_W^2 \over 4 \pi^2 v^2} M_2 \lambda x \cos \beta \sin \phi_c
f(m_{{\tilde \chi}_1}^2, \ m_{{\tilde \chi}_2}^2) \cr
& &\mbox{}+ {m_W^4 \over 2 \pi^2 v^2}
{M_2 \lambda x \cos \beta \sin \phi_c \over
(m_{{\tilde \chi}_2}^2 - m_{{\tilde \chi}_1}^2)}
\log \left ( {m_{{\tilde \chi}_2}^2 \over m_{{\tilde \chi}_1}^2} \right)
 \ , \cr
 & & \cr
\delta M_{14}^{\chi} & = &\mbox{} - {m_W^2 \over 8 \pi^2 v}
{\lambda \cos \beta \Delta_{{\tilde \chi}_1} \Delta_{\tilde \chi}
\over (m_{{\tilde \chi}_2}^2 - m_{{\tilde \chi}_1}^2)^2}
g(m_{{\tilde \chi}_1}^2, \ m_{{\tilde \chi}_2}^2) \cr
& &\mbox{} + {m_W^2 \over 4 \pi^2 v}
\lambda \cos \beta (\lambda x - M_2 \tan \beta \cos \phi_c)
f(m_{{\tilde \chi}_1}^2, \ m_{{\tilde \chi}_2}^2) \\
& &\mbox{} - {m_W^2 \lambda^2 x \cos \beta \over 8 \pi^2 v}
\log \left(
{m_{{\tilde \chi}_1}^2 m_{{\tilde \chi}_2}^2 \over \Lambda^4} \right)
- {m_W^2 \over 8 \pi^2 v}
{\lambda \cos \beta
(\lambda x \Delta_{{\tilde \chi}_1} + \Delta_{\tilde \chi}) \over
(m_{{\tilde \chi}_2}^2 - m_{{\tilde \chi}_1}^2) }
\log \left ( {m_{{\tilde \chi}_2}^2 \over m_{{\tilde \chi}_1}^2} \right)
 \ , \cr
 & & \cr
\delta M_{15}^{\chi} & = &\mbox{}
- {m_W^4 \over 4 \pi^2 v}
{M_2 \lambda \cos \beta \sin 2 \beta \sin \phi_c \Delta_{{\tilde \chi}_1}
\over (m_{{\tilde \chi}_2}^2 - m_{{\tilde \chi}_1}^2)^2}
g(m_{{\tilde \chi}_1}^2, \ m_{{\tilde \chi}_2}^2) \cr
& &\mbox{} + {m_W^2 \over 4 \pi^2 v}
M_2 \lambda \sin \beta \sin \phi_c
f(m_{{\tilde \chi}_1}^2, \ m_{{\tilde \chi}_2}^2) \cr
& &\mbox{} - {m_W^4 \over 4 \pi^2 v}
{M_2 \lambda \cos \beta \sin 2 \beta \sin \phi_c \over
(m_{{\tilde \chi}_2}^2 - m_{{\tilde \chi}_1}^2)}
\log \left ( {m_{{\tilde \chi}_2}^2 \over m_{{\tilde \chi}_1}^2} \right)
 \ , \cr
 & & \cr
\delta M_{23}^{\chi} & = &\mbox{}
{m_W^4 \over 2 \pi^2 v^2}
{M_2 \lambda x \sin \beta \sin \phi_c \Delta_{{\tilde \chi}_2} \over
(m_{{\tilde \chi}_2}^2 - m_{{\tilde \chi}_1}^2)^2}
g(m_{{\tilde \chi}_1}^2, \ m_{{\tilde \chi}_2}^2)  \cr
& &\mbox{} - {m_W^2 \over 4 \pi^2 v^2}
M_2 \lambda x \sin \beta \sin \phi_c
f(m_{{\tilde \chi}_1}^2, \ m_{{\tilde \chi}_2}^2) \cr
& &\mbox{}+ {m_W^4 \over 2 \pi^2 v^2}
{M_2 \lambda x \sin \beta \sin \phi_c \over
(m_{{\tilde \chi}_2}^2 - m_{{\tilde \chi}_1}^2)}
\log \left ( {m_{{\tilde \chi}_2}^2 \over m_{{\tilde \chi}_1}^2} \right)
 \ , \cr
 & & \cr
\delta M_{24}^{\chi} & = &\mbox{} - {m_W^2 \over 8 \pi^2 v}
{\lambda \sin \beta \Delta_{{\tilde \chi}_2} \Delta_{\tilde \chi}
\over (m_{{\tilde \chi}_2}^2 - m_{{\tilde \chi}_1}^2)^2}
g(m_{{\tilde \chi}_1}^2, \ m_{{\tilde \chi}_2}^2) \cr
& &\mbox{}+ {m_W^2 \over 4 \pi^2 v} \lambda \sin \beta
(\lambda x - M_2 \cot \beta \cos \phi_c)
f(m_{{\tilde \chi}_1}^2, \ m_{{\tilde \chi}_2}^2) \cr
& &\mbox{} -{m_W^2 \lambda^2 x \sin \beta \over 8 \pi^2 v}
\log \left( {m_{{\tilde \chi}_1}^2 m_{{\tilde \chi}_2}^2 \over \Lambda^4}
\right)
- {m_W^2 \over 8 \pi^2 v}
{\lambda \sin \beta
(\lambda x \Delta_{{\tilde \chi}_2} + \Delta_{\tilde \chi}) \over
(m_{{\tilde \chi}_2}^2 - m_{{\tilde \chi}_1}^2)}
\log \left( {m_{{\tilde \chi}_2}^2 \over m_{{\tilde \chi}_1}^2}
\right)   \ ,    \cr
 & & \cr
\delta M_{25}^{\chi} & = &\mbox{}
- {m_W^4 \over 4 \pi^2 v}
{M_2 \lambda \sin \beta \sin 2 \beta \sin \phi_c \Delta_{{\tilde \chi}_2}
\over (m_{{\tilde \chi}_2}^2 - m_{{\tilde \chi}_1}^2)^2}
g(m_{{\tilde \chi}_1}^2, \ m_{{\tilde \chi}_2}^2)  \cr
& &\mbox{} + {m_W^2 \over 4 \pi^2 v}
M_2 \lambda \cos \beta \sin \phi_c
f(m_{{\tilde \chi}_1}^2, \ m_{{\tilde \chi}_2}^2) \cr
& &\mbox{} - {m_W^4 \over 4 \pi^2 v}
{M_2 \lambda \sin \beta \sin 2 \beta \sin \phi_c \over
(m_{{\tilde \chi}_2}^2 - m_{{\tilde \chi}_1}^2)}
\log \left ( {m_{{\tilde \chi}_2}^2 \over m_{{\tilde \chi}_1}^2} \right)
 \ , \cr
 & & \cr
\delta M_{34}^{\chi} & = &\mbox{}
{m_W^2 \over 4 \pi^2 v}
{M_2 \lambda^2 x \sin \phi_c \Delta_{\tilde \chi}
\over (m_{{\tilde \chi}_2}^2 - m_{{\tilde \chi}_1}^2)^2}
g(m_{{\tilde \chi}_1}^2, \ m_{{\tilde \chi}_2}^2)  \cr
& &\mbox{} - {m_W^2 \over 4 \pi^2 v}
M_2 \lambda \sin \phi_c
f(m_{{\tilde \chi}_1}^2, \ m_{{\tilde \chi}_2}^2)
+ {m_W^2 \over 4 \pi^2 v}
{M_2 \lambda^3 x^2 \sin \phi_c \over
(m_{{\tilde \chi}_2}^2 - m_{{\tilde \chi}_1}^2)}
\log \left ( {m_{{\tilde \chi}_2}^2 \over m_{{\tilde \chi}_1}^2} \right)
 \ , \cr
 & & \cr
\delta M_{35}^{\chi} & = &\mbox{}
{m_W^4 \over 2 \pi^2 v}
{M_2^2 \lambda^2 x \sin 2 \beta \sin^2 \phi_c
\over (m_{{\tilde \chi}_2}^2 - m_{{\tilde \chi}_1}^2)^2}
g(m_{{\tilde \chi}_1}^2, \ m_{{\tilde \chi}_2}^2)  \cr
& &\mbox{} - {m_W^2 \over 4 \pi^2 v}
M_2 \lambda \cos \phi_c
f(m_{{\tilde \chi}_1}^2, \ m_{{\tilde \chi}_2}^2)
 \ , \cr
 & & \cr
\delta M_{45}^{\chi} & = &\mbox{}
- {m_W^2 \over 8 \pi^2}
{M_2 \lambda^2 \sin 2 \beta \sin \phi_c \Delta_{\tilde \chi}
\over (m_{{\tilde \chi}_2}^2 - m_{{\tilde \chi}_1}^2)^2}
g(m_{{\tilde \chi}_1}^2, \ m_{{\tilde \chi}_2}^2)  \cr
& &\mbox{} - {m_W^2 \over 8 \pi^2}
{M_2 \lambda^3 x \sin 2 \beta \sin \phi_c \over
(m_{{\tilde \chi}_2}^2 - m_{{\tilde \chi}_1}^2)}
\log \left ( {m_{{\tilde \chi}_2}^2 \over m_{{\tilde \chi}_1}^2} \right)
\ , \nonumber
\end{eqnarray}
where we define
\begin{eqnarray}
\Delta_{{\tilde \chi}_1} & = &
M_2^2 +\lambda^2 x^2 - 2 M_2 \lambda x \tan \beta \cos \phi_c \ , \cr
\Delta_{{\tilde \chi}_2} & = &
M_2^2 +\lambda^2 x^2 - 2 M_2 \lambda x \cot \beta \cos \phi_c \ , \\
\Delta_{\tilde \chi} & = &
\lambda^3 x^3 - M_2^2 \lambda x + 2 m_W^2 \lambda x
- 2 m_W^2 M_2 \sin 2 \beta \cos \phi_c \ , \nonumber
\end{eqnarray}
and
\begin{eqnarray}
 f(m_1^2, \ m_2^2) & = & {1 \over (m_2^2-m_1^2)}
\left[  m_1^2 \log {m_1^2 \over \Lambda^2} -m_2^2
\log {m_2^2 \over \Lambda^2} \right] + 1 \ , \cr
 & & \cr
 g(m_1^2,m_2^2) & = & {m_2^2 + m_1^2 \over m_1^2 - m_2^2}
        \log {m_2^2 \over m_1^2} + 2 \ .
\end{eqnarray}
At the tree level, in $M_{ij}^0$, the scalar-pseudoscalar mixing between two Higgs doublets would not occur from the Higgs potential since both $M_{13}^0$ and $M_{23}^0$ are zero.
Meanwhile none of the other elements for the scalar-pseudoscalar mixing in $M_{ij}^0$ are zero.
Assuming the nondegeneracy of the scalar quark masses, the radiative corrections due to the quark and scalar quark for the third generation yield the scalar-pseudoscalar mixing between two Higgs doublets.
Note that at the 1-loop level, on the contrary, every element of $\delta M_{ij}^{\chi}$ is nonzero; the magnitudes of these elements are proportion to $\sin \phi_c$.

It is impossible to obtain analytic expressions for the eigenvalues of $M_{ij}$ as functions of the various parameters. 
We carry out the job of obtaining the eigenvalues numerically.
The masses of the neutral Higgs bosons are then sorted such that $m_{h_i} \leqslant m_{h_j}$ for $i < j$, such that $m_{h_1}$ is the mass of the lightest neutral Higgs boson.
For the numerical calculations, we set the ranges of the relevant parameters in the Higgs potential.
The renormalization scale in the effective potential is taken to be 500 GeV.
We fix $m_t$ = 175 GeV for the top quark mass [17], $m_b$ = 4 GeV for the bottom quark mass, $m_W$ = 80.4 GeV for the charged weak gauge boson mass, and $m_Z$ = 91.1 GeV for the neutral weak gauge boson mass.
The upper bounds on $\lambda$ and $k$ are estimated as 0.87 and 0.63, respectively, by the renormalization group analysis of the NMSSM [18]. 
Thus, we set their ranges as 0 $< \lambda \leqslant$ 0.87 and 0 $< k \leqslant$ 0.63.
The ranges for other parameters are set as follows: 2 $\leqslant \tan \beta \leqslant$ 40, 0 $<$ $A_{\lambda}$, $A_k$, $x$, $m_Q$, $m_T$ (= $m_B$) $\leqslant$ 1000 GeV, 0 $<$ $A_t = A_b$ $\leqslant$ 2000 GeV, and 0 $< M_2 \leqslant$ 500 GeV.
Note that there are additional parameters at the 1-loop level, arising from $V^{\chi}$: the relative phase ($\phi_c$) between complex $M_2$ and $\lambda$, as well as $\phi$. 
We assume that they may vary from 0 to $\pi$.
We further assume that the lighter scalar quark masses of the third generation are larger than the top quark mass.

Then, we impose phenomenological constraints on the values of the parameters.
From the negative results at the LEP2 experiments on the neutral Higgs boson, research has already suggested some constraints on the parameter space of the NMSSM with explicit CP violation [15].
At the center of mass energy of LEP2, $\sqrt{s}$ = 200 GeV, the dominant production processes for the neutral Higgs boson are (i) the Higgsstrahlung process, $e^+ e^-$ $\rightarrow$ $Z^*$ $\rightarrow$ $Z h_i$ ($i$ = 1 to 5) and (ii) the Higgs-pair production process, $e^+ e^-$ $\rightarrow$ $Z^*$ $\rightarrow$ $h_i h_j$ ($i,j$ = 1 to 5, $i \neq j$).
The total cross section for producing any one of the neutral Higgs bosons in $e^+e^-$ collisions is then defined by the sum of the two processes:
$\sigma_t$ = $\sum_i \sigma_i$ + $\sum_{i j}$ $\sigma_{ij}$/2 for $i \neq j$.
Taking the discovery limit of LEP2 as 0.1 pb for $\sqrt{s}$ = 200 GeV, we exclude the parameter values if they produce $\sigma_t \geqslant$ 0.1 pb.

We introduce a useful expression in order to evaluate the effects of explicit CP violation on the neutral Higgs boson masses as in the case of the MSSM with CP violation in the Higgs potential [19].
It is the dimensionless parameter given by
\begin{equation}
\rho = 5^5 O_{11}^2 O_{21}^2 O_{31}^2 O_{41}^2 O_{51}^2  \ ,
\end{equation}
where $O_{ij}$ ($i,j$ = 1  to 5) are the elements of the orthogonal transformation matrix that diagonalizes the neutral Higgs boson mass matrix.
The range of $\rho$ is from 0 to 1 since the elements of the transformation matrix satisfy the orthogonality condition of $\sum_{j = 1}^{5} O_{j1}^2$ = 1.
If $\rho = 0$, there is no explicit CP violation in the Higgs sector of the NMSSM.
On the other hand, if $\rho = 1$, CP symmetry is maximally violated through the mixing between the scalar and pseudoscalar Higgs bosons.
The maximal CP violation that leads to $\rho$ = 1 takes place when $O_{11}^2 = O_{21}^2 = O_{31}^2 = O_{41}^2 = O_{51}^2$ = 1/5.
The elements $O_{ij}$ determine the couplings of the physical neutral Higgs bosons to the other states in the NMSSM.
Elsewhere, the chargino sector contributions, i.e., the contributions from the loops of the $W$ boson, the charged Higgs boson, and the charginos, to the neutral Higgs boson masses as well as to the mixings among them have been analyzed within the context of the MSSM with explicit CP violation [5].

Let $m_{h_1}$ denote the full 1-loop mass of the lightest neutral Higgs boson, including the contributions from the loops of the $W$ boson, the charged Higgs boson, and the charginos, as well as other loops, and $m_{h_1}^0$ denote the corresponding mass including all the loop contributions except the loops of the $W$ boson, the charged Higgs boson, and the charginos. 
Therefore,  $m_{h_1}^0$ is not the tree level mass but the 1-loop mass with the radiative corrections due to the top quark and scalar top quark loops as well as the bottom quark and scalar bottom quark loops. 
Only the chargino sector contributions are absent in it.
Then, the mass difference between $m_{h_1}$ and $m_{h_1}^0$, defined as
\[
    m_{h_1}^{\chi} = m_{h_1} - m_{h_1}^0  \ ,
\]
would evidently express the amount of the contributions from $M_{ij}^{\chi}$, and hence $V^{\chi}$, to $m_{h_1}$.

We scan the parameter space of the NMSSM using the Monte Carlo method and select points that are consistent with the above constraints.
The selected points are physically allowed in the sense that at those points the Higgs boson masses are positive and $\sigma_t$ is smaller than 0.1 pb.
Those points consist of the allowed region in the parameter space of the NMSSM.
We search in the allowed region for several points where $m_{h_1}^{\chi}$ is large and/or $\rho$ is large, i.e., where the effect of the chargino sector contributions to the smallest mass of the neutral Higgs bosons and/or the mixing among them is large.
Figure 1 shows three such points with large $m_{h_1}^{\chi}$ and/or $\rho$ as illustration. 
In Fig. 1, we plot $m_{h_1}^0$, $m_{h_1}$, $|m_{h_1}^{\chi}|/ m_{h_1}$, and $\rho$, against the CP phase $\phi_c$. 
For $|m_{h_1}^{\chi}| / m_{h_1}$ and $\rho$, we express them in \%.
The parameter values for Fig. 1(a) are $\phi$ = $\pi$/2, $\tan \beta$ = 3, $\lambda = k$ = 0.5, $A_{\lambda}$ = 800 GeV, $A_k = A_t$ = 1000 GeV, $x$ (= $m_T$) = 500 GeV, $m_Q$ = 900 GeV, and $M_2$ = 300 GeV, while the parameter values for Fig. 1(b) are $\phi$ = $\pi$/2, $\tan \beta$ = 10, $\lambda$ = 0.03, $A_{\lambda}$ = 60 GeV, $k$ = 0.4, $A_k$ = 200 GeV, $x$ = 500 GeV, $m_Q$ = $m_T$ = $A_t$ = 600 GeV, $M_2$ = 100 GeV, and for Fig. 1(c) are $\phi$ = $\pi$/2, $\tan \beta$ = 10, $\lambda$ = 0.7, $A_{\lambda}$ = 400 GeV, $k$ = 0.5, $A_k$ = 20 GeV, $x$ = 50 GeV, $m_Q$ = 500 GeV, $m_T$ = $A_t$ = 1000 GeV, $M_2$ = 100 GeV.

For the parameter values of Fig. 1(a), it is found that all of the neutral Higgs bosons $m_{h_i}$ ($i$ = 1 to 5) as well as $\sigma_t$ are roughly stable against the variation of $\phi_c$. 
They are numerically obtained as $m_{h_1}$ = 25, $m_{h_2}$ = 127, $m_{h_3}$ = 786, $m_{h_4}$ = 819, $m_{h_5}$ = 902, $m_{C^+}$ = 815 GeV, and $\sigma_t$ = 0.5 fb for $\sqrt{s}$ = 200 GeV.
One can see in Fig. 1(a) that the curve for $m_{h_1}^0 = 42$ GeV is flat.
This is because $m_{h_1}^0$ contains neither $\phi_c$ nor $M_2$.
Note that the effective potential for the contributions of the quark and scalar quark loop for the third generation is independent of both $ \phi_c$ and $M_2$.
In Fig. 1(a), $m_{h_1}^{\chi}$ is always negative for the whole range of $\phi_c$ and slightly decreasing. 
Thus, $m_{h_1}$ (GeV) is roughly stable but decreases quite slightly as $\phi_c$ increases from $0$ to $\pi$.
Hence, the slight decrease in $|m_{h_1}^{\chi}| / m_{h_1}$ for larger $\phi_c$.
It is observed that $\rho$ increases as $\phi_c$ goes from 0 to $\pi$.
This behavior of $\rho$ arises from destructive interferences between the chargino sector contributions and the other ones.

Figure 1(b) displays a relatively large $m_{h_1}$. 
We have $m_{h_1}$ = 94, $m_{h_2}$ = 98, $m_{h_3}$ = 128, $m_{h_4}$ = 346, $m_{h_5}$ = 346, $m_{C^+}$ = 124 GeV, and $\sigma_t$ = 36 fb. 
For the parameter values of Fig. 1(b), the neutral Higgs boson masses as well as $\sigma_t$ are almost stable against $\phi_c$, similar to Fig. 1(a).
One sees that $m_{h_1}^0$ and $m_{h_1}$ are nearly equal, and large as compared to the case of Fig. 1(a). 
Thus, $m_{h_1}^{\chi}$ for the parameter values of Fig. 1(b) is very small.
It is positive for smaller $\phi_c$ and negative for larger $\phi_c$.
There is a crossover at about $\phi_c \approx 2\pi/3$.
On the other hand, $\rho$ shows interesting behavior.
It varies quite abruptly; it drops from 100\% at $\phi_c = 0$ down to zero at about $\phi_c \approx 0.35\pi$, and then increases as large as 80\%.
However, it is difficult to analyze why $\rho$ varies so widely mainly because of the nature of the Higgs sector of the NMSSM. The complicated form of the $5\times5$ mass matrix of the neutral Higgs bosons prohibits us to trace analytically down the behavior of $\rho$.

For the parameter values of Fig. 1(c), $m_{h_1}^{\chi}$ is always positive but decreasing from 13 GeV to 5 GeV as $\phi_c$ goes from 0 to $\pi$.
Both $m_{h_1}^0$ and $m_{h_1}$ in Fig. 1(c) are comparable in size to the case of Fig. 1(a).
Approximately, we have $m_{h_1}$ = 38, $m_{h_2}$ = 52, $m_{h_3}$ = 124, $m_{h_4}$ = 392, $m_{h_5}$ = 403, $m_{C^+}$ = 364 GeV, and $\sigma_t$ = 10 fb in Fig. 1(c). For $m_{h_1}^0$, we have about 25 GeV.
The behavior of $\rho$ for the parameter values of Fig. 1(c) is as varied as that of the case of Fig. 1(b). 
We note that the three sets of parameter values used in Fig. 1 are consistent with the LEP2 data, as indicated by the size of the total cross sections for Higgs production: None of $\sigma_t$ is larger than 0.1 pb, the discovery limit of LEP2.

Now, let us study the behavior of $m_{h_1}^{\chi}$ in detail.
It comes from the 1-loop effective potential $V^{\chi}$, which includes the effects of the chargino mass splitting and possesses the CP phase $\phi_c$, which in turn depends on the SU(2) gaugino mass $M_2$.
It represents the contributions of the loops of the $W$ boson, charged Higgs boson, and the charginos.
Thus, the dependence of $m_{h_1}^{\chi}$ on $\phi_c$ and $M_2$ would give us a measure of the effect of explicit CP violation in the NMSSM.

Figure 2 shows the dependence of $m_{h_1}^{\chi}$ on them, plotted against the ratio of the two chargino masses $m_{{\tilde \chi}_2}/m_{{\tilde \chi}_1}$.
In Fig. 2 there are three sets of curves, and each set consists of three curves corresponding to three different values of $\phi_c$: $\phi_c = 0$ (solid curve), $\pi/2$ (dashed curve), and $\pi$ (dotted curve). 
The extent of each curve corresponds to the variation of $M_2$: We vary $M_2$ from 1 to 500 GeV. For the rest of relevant parameters, we set their values as: $\phi$ = $\pi$/2, $\tan \beta$ = 3, $\lambda = k$ = 0.5, $A_{\lambda}$ = 800 GeV, $A_k = A_t$ = 1000 GeV, $x$ (= $m_T$) = 500 GeV, and $m_Q$ = 900 GeV for the set of curves of feature (A1); $\phi$ = $\pi$/2, $\tan \beta$ = 10, $\lambda$ = 0.03, $A_{\lambda}$ = 60 GeV, $k$ = 0.4, $A_k$ = 200 GeV, $x$ = 500 GeV, and $m_Q$ = $m_T$ = $A_t$ = 600 GeV for the set of curves of feature (A2), and $\phi$ = $\pi$/2, $\tan \beta$ = 10, $\lambda$ = 0.7, $A_{\lambda}$ = 400 GeV, $k$ = 0.5, $A_k$ = 20 GeV, $x$ = 50 GeV, $m_Q$ = 500 GeV, and $m_T$ = $A_t$ = 1000 GeV for the set of curves of feature (A3).
In short, the parameter values for the three sets of curves of features (A1), (A2), and (A3) are the same as Figs. 1(a), 1(b), and 1(c), respectively, except for $\phi_c$ and $M_2$.

When both $\phi_c$ and $M_2$ approach zero as the charginos becomes degenerate in mass, $m_{{\tilde \chi}_2}/m_{{\tilde \chi}_1}$ approaches 1. 
Still in this case, if $\phi_c = 0$, CP violation in the Higgs sector is possible from $V^0 + V^t + V^b$ since $\phi$ is fixed as $\pi/2$.
For the parameter values of feature (A1), $m_{h_1}^{\chi}$ is shown to be relatively independent of the variation in $\phi_c$, and found to increase as $M_2$ increases.
Nevertheless, $m_{h_1}^{\chi}$ is always negative from $- 24$ to about $- 16$ GeV for the whole range of $M_2$. 
The ratio of the chargino masses decreases from about 15 to 2 as $M_2$ increases from 1 to 500 GeV for the parameter values of feature (A1).
On the contrary, for the parameter values of both features (A2) and (A3), the ratio of the chargino masses increases as $M_2$ increases from 1 to 500 GeV: For those of feature (A2), the ratio of the chargino masses increases from 1 to as large as 30, and for those of feature (A3), from 1 to 15.

The variation in $m_{h_1}^{\chi}$ for the parameter values of feature (A2) is rather small: $\pm 2$ GeV for the whole ranges of $0 < M_2 \leqslant$ 500 GeV and of $0 \leqslant \phi_c \leqslant \pi$.
In contrast, the variation in $m_{h_1}^{\chi}$ for the parameter values of feature (A3) is very large: it can be as large as 16 GeV. Therefore, the overall view of Fig. 2 implies that $m_{h_1}^{\chi}$ is not negligible at all. 
Our analysis shows that for some particular values of the relevant parameters $m_{h_1}^{\chi}$ may have any value between $- 24$ and 16 GeV.
Thus, it may play a quite crucial role in the analysis of the explicit CP violation scenario of the NMSSM.

\section{NEUTRAL HIGGS PRODUCTION IN $e^+e^-$ COLLISIONS}

We know that the unsuccessful result at LEP2 for the Higgs search does not exclude the possibility of the existence of a massless neutral Higgs boson in the NMSSM with explicit CP violation.
In this section we investigate the detectability of a neutral Higgs boson in $e^+e^-$ collisions with much higher center-of-mass energy, such as at the future $e^+e^-$ linear collider with $\sqrt{s}$ = 500 (LC500) GeV and 1000 (LC1000) GeV. We consider the case in which the lightest neutral Higgs boson of the NMSSM is composed dominantly of the singlet field.

Let us first derive an upper bound on the mass of the lightest neutral Higgs boson in the NMSSM with explicit CP violation, where the Higgs singlet field contribute dominantly.
The square of the radiatively corrected upper bound on the lightest neutral Higgs boson mass,
$m_{h_1, \ {\rm max}}^2$ is given as 
\begin{eqnarray}
& & m_Z^2 + (\lambda^2 v^2 - m_Z^2) \sin^2 2 \beta \cr
& &\mbox{} + {3 m_t^4 \over 8 \pi^2 v^2}
{(\lambda x \cot \beta \Delta_{{\tilde t}_1} + A_t \Delta_{{\tilde t}_2})^2
\over (m_{{\tilde t}_2}^2 - m_{{\tilde t}_1}^2)^2}
g(m_{{\tilde t}_1}^2, \ m_{{\tilde t}_2}^2) \cr
     &  & \mbox{} + {3 m_t^4 \over 4 \pi^2 v^2}
{(\lambda x \cot \beta \Delta_{{\tilde t}_1} + A_t \Delta_{{\tilde t}_2})
\over (m_{{\tilde t}_2}^2 - m_{{\tilde t}_1}^2)}
\log \left({m_{{\tilde t}_2}^2 \over m_{{\tilde t}_1}^2}\right)
+ {3 m_t^4 \over 8 \pi^2 v^2}
\log({m_{{\tilde t}_1}^2 m_{{\tilde t}_2}^2 \over m_t^4}) \cr
& &\mbox{} + {3 m_b^4 \over 8 \pi^2 v^2}
{(\lambda x \tan \beta \Delta_{{\tilde b}_1} + A_t \Delta_{{\tilde b}_2})^2
\over (m_{{\tilde b}_2}^2 - m_{{\tilde b}_1}^2)^2}
g(m_{{\tilde b}_1}^2, \ m_{{\tilde b}_2}^2) \cr
     &  & \mbox{} + {3 m_b^4 \over 4 \pi^2 v^2}
{(\lambda x \tan \beta \Delta_{{\tilde b}_1} + A_t \Delta_{{\tilde b}_2})
\over (m_{{\tilde b}_2}^2 - m_{{\tilde b}_1}^2)}
\log \left({m_{{\tilde b}_2}^2 \over m_{{\tilde b}_1}^2}\right)
+ {3 m_b^4 \over 8 \pi^2 v^2}
\log({m_{{\tilde b}_1}^2 m_{{\tilde b}_2}^2 \over m_b^4})    \cr
     &  & \mbox{} - {m_W^4 \over 4 \pi^2 v^2}
{(\cos^2 \beta \Delta _{{\tilde \chi}_1}
+ \sin^2 \beta \Delta _{{\tilde \chi}_2})^2
\over (m_{{\tilde \chi}_2}^2 - m_{{\tilde \chi}_1}^2)^2}
g(m_{{\tilde \chi}_1}^2, \ m_{{\tilde \chi}_2}^2) 
+ {m_W^4 \over 8 \pi^2 v^2} \log \left ({m_W^6 m_{C^+}^2 \over
m_{{\tilde \chi}_1}^4 m_{{\tilde \chi}_2}^4} \right )   \cr
    &  &\mbox{} - {m_W^4 \over 2 \pi^2 v^2}
{\log ({m_{{\tilde \chi}_2}^2 / m_{{\tilde \chi}_1}^2})
\over (m_{{\tilde \chi}_2}^2 - m_{{\tilde \chi}_1}^2)}
\{\cos^4 \beta \Delta_{{\tilde \chi}_1} + \sin^4 \beta \Delta_{{\tilde \chi}_2}
+ \sin^2 2 \beta (\Delta_{{\tilde \chi}_1} + \Delta_{{\tilde \chi}_2}) \} \ ,
\end{eqnarray}
with
\begin{eqnarray}
 \Delta_{{\tilde t}_1} & = & A_t \cos \phi + \lambda x \cot \beta  \  , \cr
 \Delta_{{\tilde t}_2} & = & A_t + \lambda x \cot \beta \cos \phi \ , \cr
 \Delta_{{\tilde b}_1} & = & A_t \cos \phi + \lambda x \tan \beta  \  , \\
 \Delta_{{\tilde b}_2} & = & A_t + \lambda x \tan \beta \cos \phi \ . \nonumber
\end{eqnarray}
Note that $\Delta_{{\tilde \chi}_1}$ and $\Delta_{{\tilde \chi}_2}$ already appear in the expression of $\delta M_{ij}^{\chi}$.

In the limit of $\cos \phi$ = $\cos \phi_c$ = 1, the above upper bound on the lightest neutral Higgs boson mass reduces to the one obtained with CP conservation in the neutral Higgs sector.
The first two terms appearing in the expression for $m_{h_1, {\rm max}}^2$ are obtained from the tree level Higgs potential. They do not have any CP phase.
The maximum value of $\lambda$ determines the tree level upper bound on the lightest neutral Higgs boson mass.
The remaining terms in $m_{h_1, {\rm max}}^2$ come from radiative corrections arising from the quark and scalar quark loops for the third generation, as well as the loops of $W$ boson, the charged Higgs boson, and the charginos.

In terms of $m_{h_1, {\rm max}}$ and $m_{h_1}$, it is useful to express the upper bound of the other neutral Higgs boson masses as [20]
\begin{eqnarray}
m_{h_2}^2 & \leqslant & m_{h_2, \ {\rm max}}^2 = { m_{h_1, \ {\rm max}}^2 - R_1^2 m_{h_1}^2 \over 1 - R_1^2 } \  , \cr 
& & \cr 
m_{h_3}^2 & \leqslant & m_{h_3, \ {\rm max}}^2 = { m_{h_1, \ {\rm max}}^2 - (\sum_{i = 1}^2 R_i^2) m_{h_1}^2 \over 1 - (\sum_{i = 1}^2 R_i^2) }  \ , \cr 
& & \cr 
m_{h_4}^2 & \leqslant & m_{h_4, \ {\rm max}}^2 = { m_{h_1, \ {\rm max}}^2 - (\sum_{i = 1}^3 R_i^2) m_{h_1}^2 \over 1 - (\sum_{i = 1}^3 R_i^2)}  \ , \\
& & \cr 
m_{h_5}^2 & \leqslant & m_{h_5, \ {\rm max}}^2 = { m_{h_1, \ {\rm max}}^2 - (\sum_{i = 1}^4 R_i^2) m_{h_1}^2 \over 1 - (\sum_{i = 1}^4 R_i^2) }  \ , \nonumber
\end{eqnarray}
where $R_i$ are defined as $R_i = (O_{i1} \cos \beta + O_{i2} \sin \beta)$ and satisfy the sum rule of $\sum_{i = 1}^5 R_i^2$ = 1.

It is known that if the center of mass energy of the colliding electrons and positrons is larger than $E_{\rm T} = m_Z + m_{h_i}$ ($i$ = 1 to 5), the Higgsstrahlung process $e^+ e^- \rightarrow Z h_i$ ($i$ = 1 to 5) is viable for the Higgs production.
The cross section $\sigma_i$ ($i$ = 1 to 5) for $h_i$ production only through the Higgsstrahlung process is known to be related to that for the Higgs production in the standard model, $\sigma_{\rm SM}$, by
\begin{eqnarray}
\sigma_j (m_{h_j}) & = &
\sigma_{\rm SM} (m_{h_j}) R_j^2 \ {\rm for} \ j = 1 \ {\rm to} \ 4 \cr
\sigma_5 (m_{h_5}) & = &
\sigma_{\rm SM} (m_{h_5}) \left (1 - \sum_{j = 1}^4 R_j^2 \right )  \ .
\end{eqnarray}
One can derive the parameter independent lower bound of $\sigma_i$ from the fact that $\sigma_i (m_{{h_i},{\rm max}})$ $\leqslant$ $\sigma_i$ ($m_{h_i}$).

In order to be systematic we assume that any of the five neutral Higgs bosons may be produced via the Higgsstrahlung process if kinematically allowed.
For each of them, we calculate the cross section for their productions.
Let us denotethe five cross sections as $\sigma_i$ ($i$ = 1 to 5); their dependence on the relevant parameters may collectively be expressed as
\begin{eqnarray}
\sigma_1 & = & \sigma_1 (R_1,R_2,R_3,R_4,m_{h_1}) \ , \cr
\sigma_k & = & \sigma_k (R_1,R_2,R_3,R_4,m_{h_j,{\rm max}}) 
\ (k=2 \ {\rm to} \ 5) \ .
\end{eqnarray}
By varying $R_j$ ($j$ = 1 to 4) and $m_{h_1}$, one can obtain the maximum values for each cross sections, which would provide us a criterion on whether it is possible to detect any of them at the future $e^+e^-$ colliders.

For given $R_j$ ($j$ = 1 to 4) and $m_{h_1}$, we may choose the largest one among the five cross sections: Let it be defined as
\begin{equation}
\sigma_0 = {\rm max} \{ \sigma_1,\sigma_2,\sigma_3,\sigma_4,\sigma_5 \} \ . 
\end{equation}
Now, we take the minimum of $\sigma_0$ by varying $R_j$ ($j$ = 1 to 4) and $m_{h_1}$. 
This would tell us whether at least one of the five neutral Higgs bosons might be detected.
Thus, we search $\sigma_0$ for 0 $\leqslant$ $R_j$ ($j$ = 1 to 4) $\leqslant$ 1 and for 0 $<$ $m_{h_1}$ $\leqslant$ $m_{h_1, \ {\rm max}}$, and set up its minimum value in order to examine the possibility of detecting at future $e^+e^-$ colliders at least one of the five neutral Higgs bosons of the NMSSM, at 1-loop level with explicit CP violation.

In Fig. 3(a) we plot $\sigma_0$, as a function of $R_1^2$ for $\sqrt{s}$ = 500 (LC500) and 1000 (LC1000) GeV. For comparison, we also calculate the corresponding quantity for the case where the CP violation is absent. 
The solid curve is the result for the case of explicit CP violation, while the dashed curve is the result with no CP violation. 
From Fig. 3(a), one sees that for the case of explicit CP violation scenario, $\sigma_0$ can be as small as about 10 fb for $\sqrt{s}$ = 500 GeV, and 2.5 fb for $\sqrt{s}$ = 1000 GeV. 
On the other hand, for the case with no CP violation, the smallest value of $\sigma_0$ is about 16 for $\sqrt{s}$ = 500 GeV and 3.0 fb for $\sqrt{s}$ = 1000 GeV. 
By assuming an efficiency of 50$\%$, the integrated luminosity at LC500 for 50 events would be required to be about 10 fb$^{-1}$ in order to detect at least one of the five neutral Higgs bosons of the NMSSM at future $e^+e^-$ colliders if the CP symmetry is broken explicitly in the model. 
If there is no explicit CP violation in the NMSSM, the corresponding integrated luminosity needs to be 6.25 fb$^{-1}$.

In the literature, the worst case has been considered [21] when there is no explicit CP violation in the NMSSM: It is the case in which all the three neutral scalar Higgs bosons are degenerate in mass and $R_1^2 = R_2^2 = R_3^2 = 1/3$. 
For our study, with explicit CP violation, the worst case would be $R_i^2= 1/5$ ($i$ = 1 to 5). 
The five neutral Higgs bosons in our study may not be degenerate in mass.
In Fig. 3(b), we plot $\sigma$ for $R_i^2= 1/5$ ($i$ = 1 to 5).
It is found that all of $\sigma_i$ ($i$ = 2 to 5) except for $\sigma_1$ increase as $m_{h_1}$ increases.

\section{CONCLUSIONS}

We have investigated the neutral Higgs sector in the NMSSM with explicit CP violation at the 1-loop level by using the effective potential method.
For the radiatively corrected masses of the neutral Higgs bosons, we include not only the contributions of the quark and scalar quark for the third generation, but also the contributions of $W$ boson, the charged Higgs boson, and the charginos.
These additional contributions are found to be very crucial for the analysis of the neutral Higgs sector of the NMSSM, especially when the CP symmetry is broken explicitly.

There are several complex phases in the NMSSM with explicit CP violation:
At the NMSSM, three phases appear in the tree level Higgs potential and in the masses of the scalar top quarks and the scalar bottom quarks. 
We have assumed that these three phases are equal.
At the 1-loop level, besides these three phases, one additional CP phase arises from the 1-loop effective potential due to the contributions of the $W$ boson, the charged Higgs boson, and the charginos.

We have found that the mass of the lightest neutral Higgs boson may either increase by 16 GeV or decrease by about 24 GeV, due to the contributions of $W$ boson, the charged Higgs boson, and the charginos. 
This amount of variation in mass is quite significant to the relatively light $m_{h_1}$ (25-40 GeV).
Also the mixings among the neutral Higgs bosons are significantly affected both by the CP phase and the contributions arising from the contributions of the $W$ boson, the charged Higgs boson, and the charginos.

The production cross sections for the neutral Higgs bosons in the NMSSM with explicit CP violation are calculated via the Higgsstrahlung processes, $e^+ e^- \rightarrow Z h_i$ ($i$ = 1 to 5), when the lightest neutral Higgs boson is composed dominantly of the Higgs singlet field.
The minimum of the production cross section for at least one of the five neutral Higgs bosons in the NMSSM is calculated to be about 10 and 2.5 fb for $\sqrt{s}$ = 500 and 1000 GeV, respectively.
These values are about one-third smaller than the corresponding numbers obtained by assuming CP conservation in the NMSSM Higgs sector.

\vskip 0.3 in

\noindent
{\large {\bf ACKNOWLEDGMENTS}}

This work was supported by a grant from Kyungpook National University.
This research is partly supported through the Science Research Center Program
by the Korea Science and Engineering Foundation.

\vskip 0.3 in


\vfil\eject

{\large {\bf FIGURE CAPTIONS}}
\vskip 0.3 in
\noindent
FIG. 1. (a) : The plots of $m_{h_1}^0$, $m_{h_1}$, $|m_{h_1}^{\chi}|/ m_{h_1}$, and $\rho$ as functions of the CP phase $\phi_c$.
For $m_{h_1}^0$ and $m_{h_1}$, we express them in GeV, and for $|m_{h_1}^{\chi}| / m_{h_1}$ and $\rho$, in \%.
The values of other relevant parameters are $\phi$ = $\pi$/2, $\tan \beta$ = 3,
$\lambda = k$ = 0.5, $A_{\lambda}$ = 800 GeV, $A_k = A_t$
= 1000 GeV, $x$ (= $m_T$) = 500 GeV, $m_Q$ = 900 GeV, and $M_2$ = 300 GeV.

\vskip 0.3 in
\noindent
FIG. 1. (b) : The same plots as in FIG. 1. (a), except for different parameter values $\phi$ = $\pi$/2, $\tan \beta$ = 10, $\lambda$ = 0.03, $A_{\lambda}$ = 60 GeV, $k$ = 0.4, $A_k$ = 200 GeV, $x$ = 500 GeV, $m_Q$ (= $m_T$ = $A_t$) = 600 GeV, $M_2$ = 100 GeV.

\vskip 0.2 in
\noindent
FIG. 1. (c) : The same plots as in FIG. 1. (a) or (b), except for different parameter values $\phi$ = $\pi$/2, $\tan \beta$ = 10, $\lambda$ = 0.7, $A_{\lambda}$ = 400 GeV, $k$ = 0.5, $A_k$ = 20 GeV, $x$ = 50 GeV, $m_Q$ = 500 GeV, $m_T$ (= $A_t$) = 1000 GeV, $M_2$ = 100 GeV.

\vskip 0.2in
\noindent
FIG. 2. : The plot of $m_{h_1}^{\chi}$ as a function of $m_{{\tilde \chi}_2}/m_{{\tilde \chi}_1}$, the ratio of the two chargino masses. 
The parameter values of the set of curves of feature (A1) are $\phi$ = $\pi$/2, $\tan \beta$ = 3, $\lambda = k$ = 0.5, $A_{\lambda}$ = 800 GeV, $A_k = A_t$ = 1000 GeV, $x$ (= $m_T$) = 500 GeV, and $m_Q$ = 900 GeV.
Those of feature (A2) are $\phi$ = $\pi$/2, $\tan \beta$ = 10, $\lambda$ = 0.03, $A_{\lambda}$ = 60 GeV, $k$ = 0.4, $A_k$ = 200 GeV, $x$ = 500 GeV, and $m_Q$ = $m_T$ = $A_t$ = 600 GeV, and those of feature (A3) are $\phi$ = $\pi$/2, $\tan \beta$ = 10, $\lambda$ = 0.7, $A_{\lambda}$ = 400 GeV, $k$ = 0.5, $A_k$ = 20 GeV, $x$ = 50 GeV, $m_Q$ = 500 GeV, and $m_T$ = $A_t$ = 1000 GeV.
Each set consists of three curves corresponding to three different values of $\phi_c$: $\phi_c = 0$ (solid curve), $\pi/2$ (dashed curve), and $\pi$ (dotted curve).
The extent of each curve corresponds to the variation of $M_2$: We vary $M_2$ from 1 to 500 GeV.

\vskip 0.2 in
\noindent
FIG. 3. (a) : The minimum production cross section for at least one of the neutral Higgs boson, $\sigma_0$, is plotted as a function of $R_1^2$, for $\sqrt{s}$ = 500 (LC500) and 1000 (LC1000) GeV.
The solid curve corresponds to the case of the explicit CP violation while the dashed curve to no CP violation.

\vskip 0.2 in
\noindent
FIG. 3. (b) : The production cross sections for each of the five neutral Higgs bosons in the NMSSM, $\sigma_i$ ($i$=1 to 5), are plotted as functions of $m_{h_1}$, for $\sqrt{s}$ = 500 GeV.

\vfil\eject

\setcounter{figure}{0}
\def\figurename{}{}%
\renewcommand\thefigure{FIG. 1. (a)}
\begin{figure}[t]
\epsfxsize=13cm
\hspace*{2.cm}
\epsffile{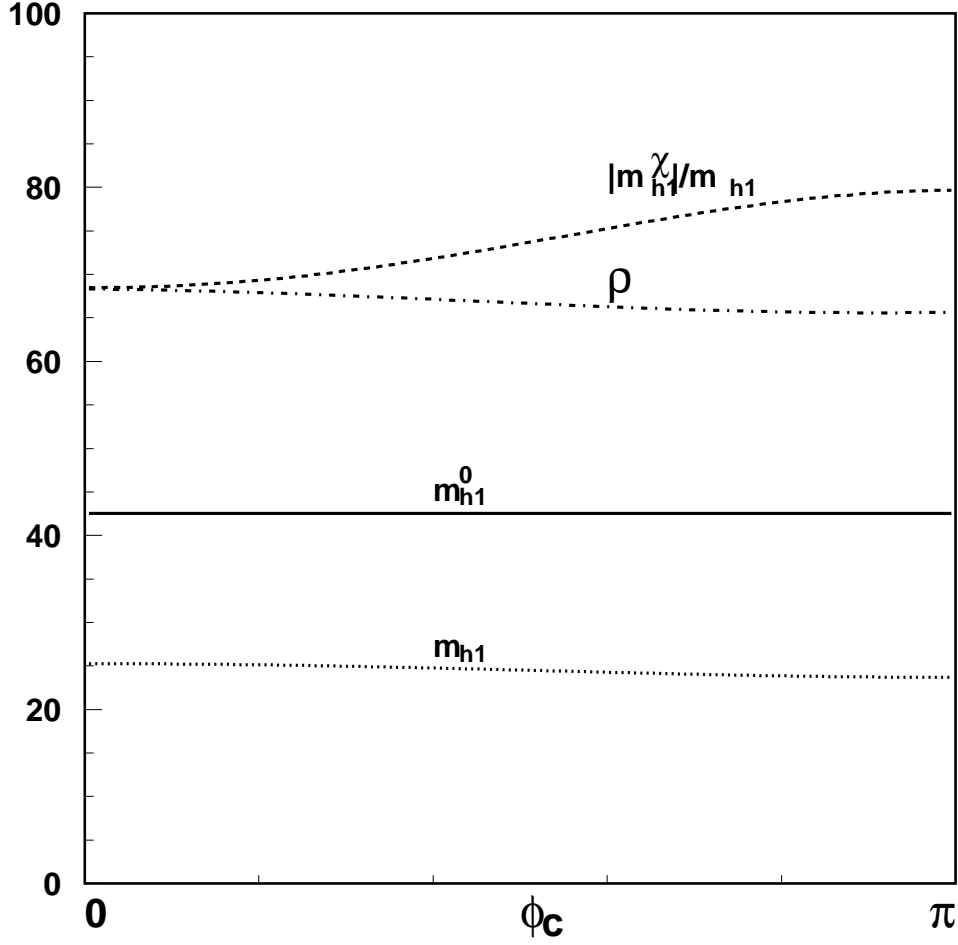}
\caption[plot]{The plots of $m_{h_1}^0$, $m_{h_1}$, $|m_{h_1}^{\chi}|/ m_{h_1}$, and $\rho$ as functions of the CP phase $\phi_c$.
For $m_{h_1}^0$ and $m_{h_1}$, we express them in GeV, and for $|m_{h_1}^{\chi}| / m_{h_1}$ and $\rho$, in \%.
The values of other relevant parameters are $\phi$ = $\pi$/2, $\tan \beta$ = 3,
$\lambda = k$ = 0.5, $A_{\lambda}$ = 800 GeV, $A_k = A_t$
= 1000 GeV, $x$ (= $m_T$) = 500 GeV, $m_Q$ = 900 GeV, and $M_2$ = 300 GeV.}
\end{figure}

\renewcommand\thefigure{FIG. 1. (b)}
\begin{figure}[t]
\epsfxsize=13cm
\hspace*{2.cm}
\epsffile{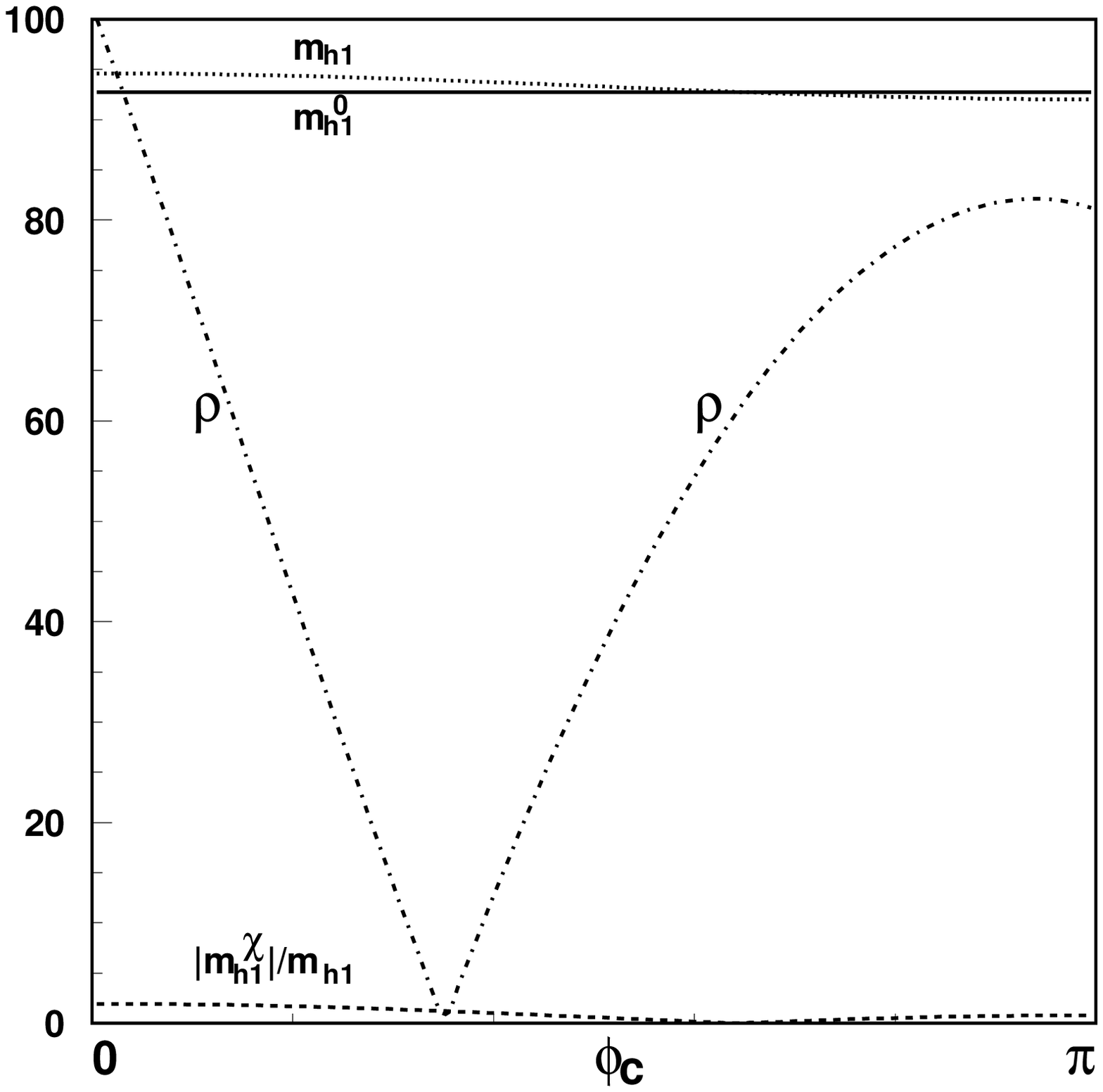}
\caption[plot]{The same plots as in FIG. 1. (a), except for different parameter values $\phi$ = $\pi$/2, $\tan \beta$ = 10, $\lambda$ = 0.03, $A_{\lambda}$ = 60 GeV, $k$ = 0.4, $A_k$ = 200 GeV, $x$ = 500 GeV, $m_Q$ (= $m_T$ = $A_t$) = 600 GeV, $M_2$ = 100 GeV.}
\end{figure}

\renewcommand\thefigure{FIG. 1. (c)}
\begin{figure}[t]
\epsfxsize=13cm
\hspace*{2.cm}
\epsffile{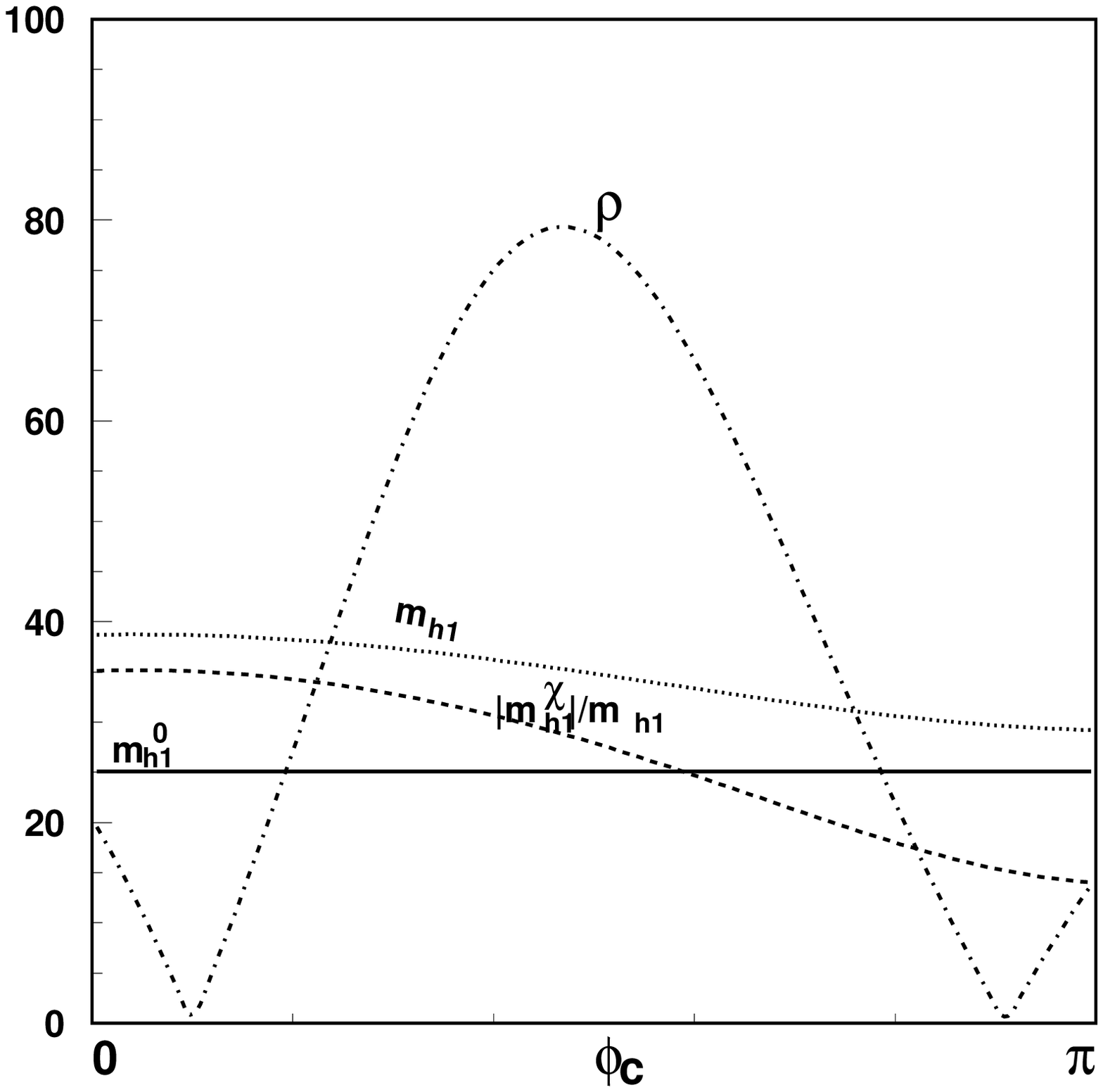}
\caption[plot]{The same plots as in FIG. 1. (a) or (b), except for different parameter values $\phi$ = $\pi$/2, $\tan \beta$ = 10, $\lambda$ = 0.7, $A_{\lambda}$ = 400 GeV, $k$ = 0.5, $A_k$ = 20 GeV, $x$ = 50 GeV, $m_Q$ = 500 GeV, $m_T$ (= $A_t$) = 1000 GeV, $M_2$ = 100 GeV.}
\end{figure}

\renewcommand\thefigure{FIG. 2.}
\begin{figure}[t]
\epsfxsize=13cm
\hspace*{2.cm}
\epsffile{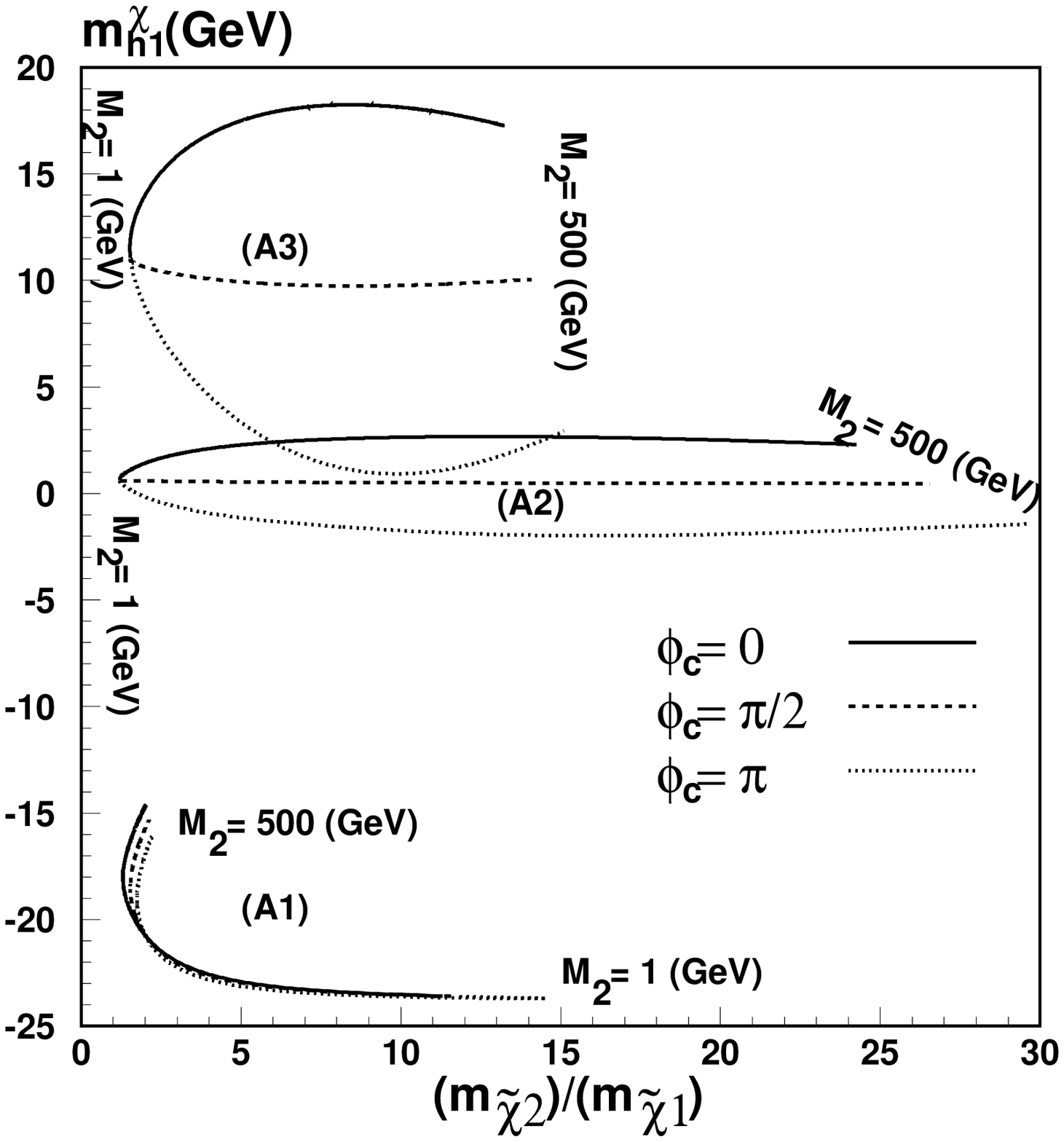}
\caption[plot]{ The plot of $m_{h_1}^{\chi}$ as a function of $m_{{\tilde \chi}_2}/m_{{\tilde \chi}_1}$, the ratio of the two chargino masses. 
The parameter values of the set of curves of feature (A1) are $\phi$ = $\pi$/2, $\tan \beta$ = 3, $\lambda = k$ = 0.5, $A_{\lambda}$ = 800 GeV, $A_k = A_t$ = 1000 GeV, $x$ (= $m_T$) = 500 GeV, and $m_Q$ = 900 GeV.
Those of feature (A2) are $\phi$ = $\pi$/2, $\tan \beta$ = 10, $\lambda$ = 0.03, $A_{\lambda}$ = 60 GeV, $k$ = 0.4, $A_k$ = 200 GeV, $x$ = 500 GeV, and $m_Q$ = $m_T$ = $A_t$ = 600 GeV, and those of feature (A3) are $\phi$ = $\pi$/2, $\tan \beta$ = 10, $\lambda$ = 0.7, $A_{\lambda}$ = 400 GeV, $k$ = 0.5, $A_k$ = 20 GeV, $x$ = 50 GeV, $m_Q$ = 500 GeV, and $m_T$ = $A_t$ = 1000 GeV.
Each set consists of three curves corresponding to three different values of $\phi_c$: $\phi_c = 0$ (solid curve), $\pi/2$ (dashed curve), and $\pi$ (dotted curve).
The extent of each curve corresponds to the variation of $M_2$: We vary $M_2$ from 1 to 500 GeV.}
\end{figure}

\renewcommand\thefigure{FIG. 3. (a)}
\begin{figure}[t]
\epsfxsize=13cm \hspace*{2.cm} \epsffile{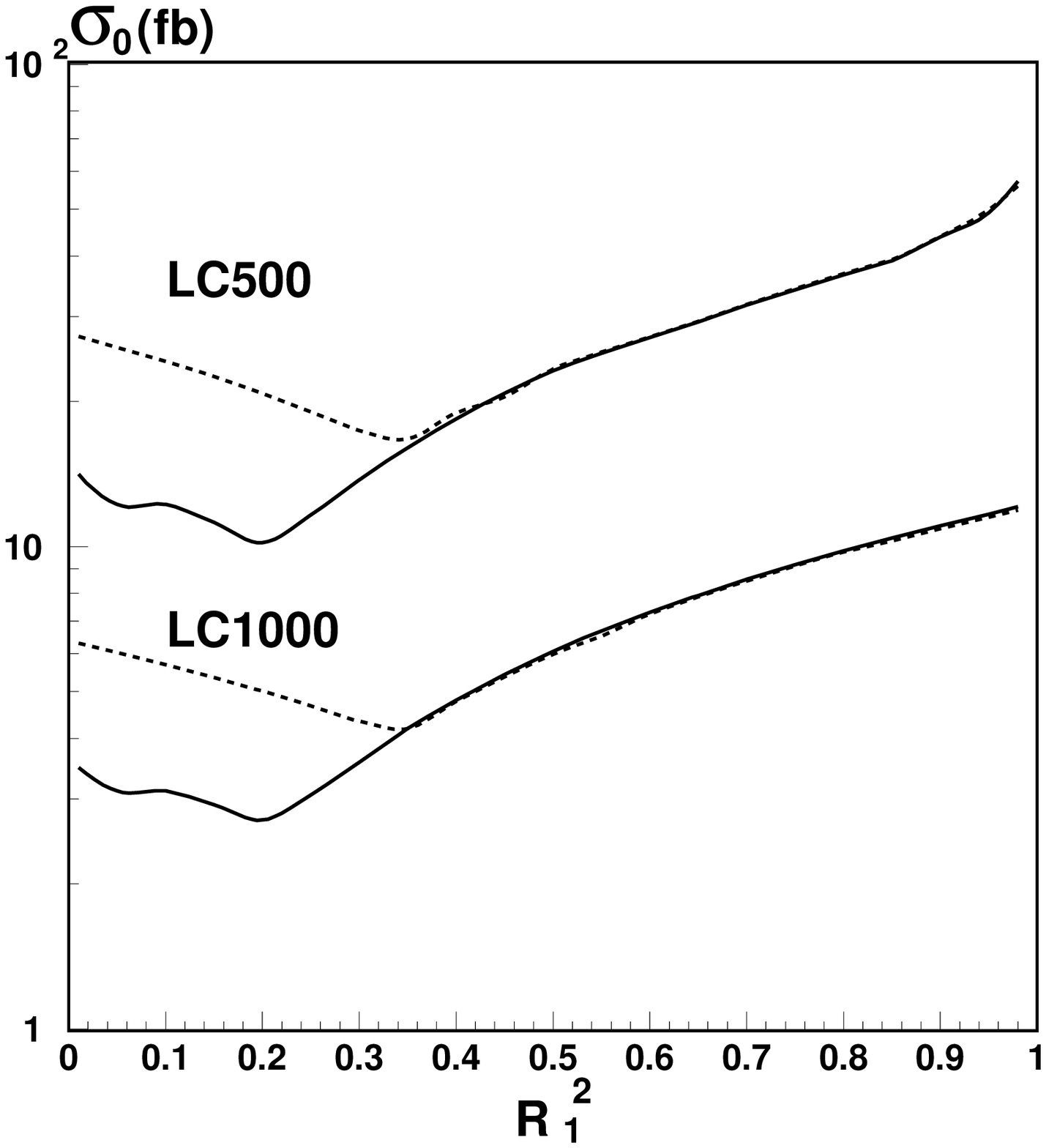}
\caption[plot]{The minimum production cross section for at least one of the neutral Higgs boson, $\sigma_0$, is plotted as a function of $R_1^2$, for $\sqrt{s}$ = 500 (LC500) and 1000 (LC1000) GeV.
The solid curve corresponds to the case of the explicit CP violation while the dashed curve to no CP violation.}
\end{figure}

\renewcommand\thefigure{FIG. 3. (b)}
\begin{figure}[t]
\epsfxsize=13cm \hspace*{2.cm} \epsffile{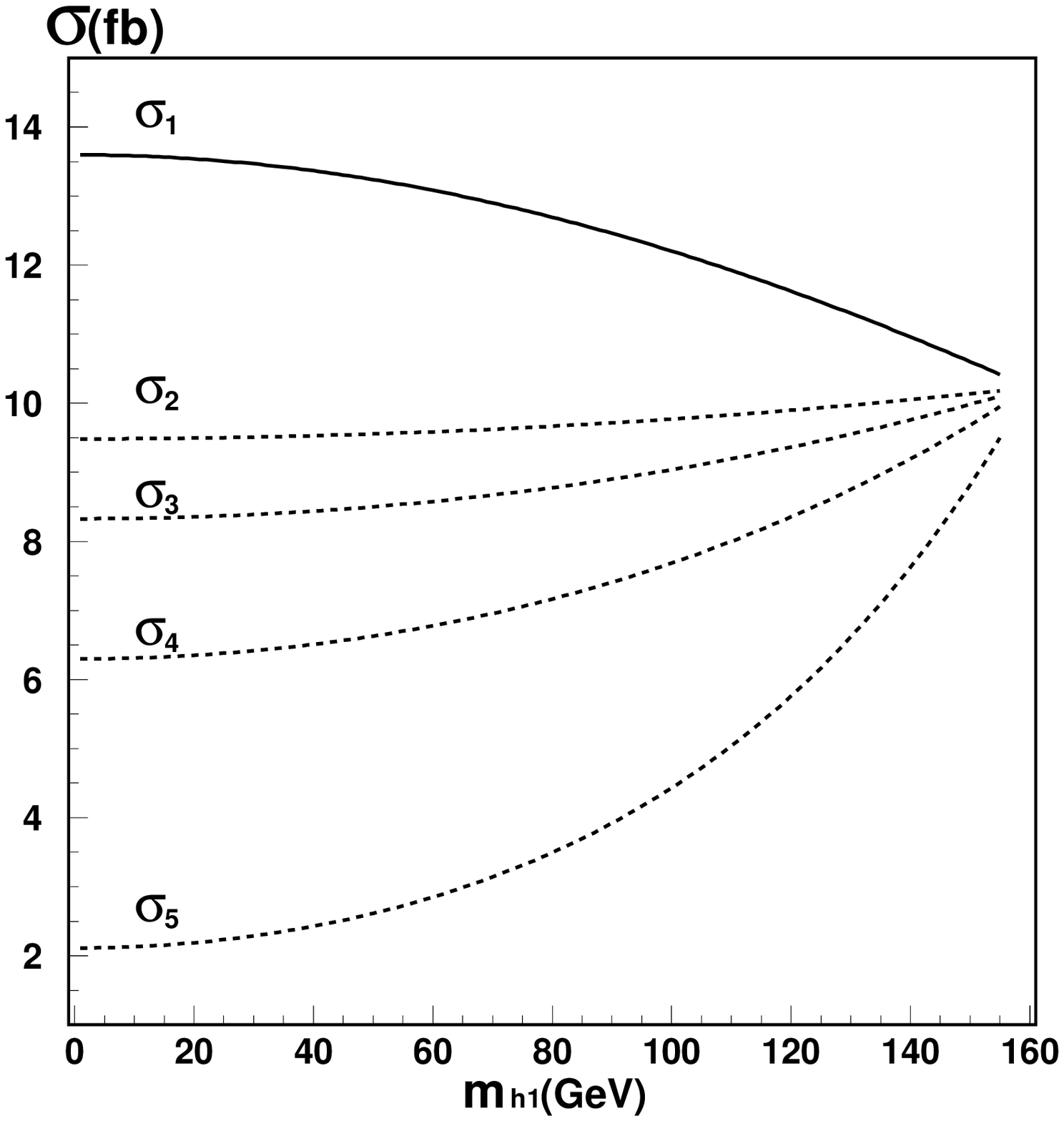}
\caption[plot]{The production cross sections for each of the five neutral Higgs bosons in the NMSSM, $\sigma_i$ ($i$=1 to 5), are plotted as functions of $m_{h_1}$, for $\sqrt{s}$ = 500 GeV.}
\end{figure}
\end{document}